\documentclass[]{spie}  

\usepackage{amsmath,amsfonts,amssymb}
\usepackage{graphicx}
\usepackage{siunitx}
\usepackage[colorlinks=true, allcolors=blue]{hyperref}

\title{The Space Coronagraph Optical Bench (SCoOB): 5. End-to-end simulations of polarization aberrations}

\author[a]{Ramya M Anche}
\author[a]{Kyle J. Van Gorkom}
\author[a,b]{Jaren N. Ashcraft}
\author[a]{Ewan Douglas}
\author[a,b]{Emory L Jenkins}
\author[c]{Sebastiaan Y. Haffert}
\author[d]{Maxwell A. Millar-Blanchaer}

\affil[a]{Steward Observatory, University of Arizona, 933N Cherry Avenue, Tucson, Arizona, 85721, USA}
\affil[b]{James C. Wyant College of Optical Sciences, University of Arizona, 933N Cherry Avenue, Tucson, Arizona, 85721, USA}
\affil[c]{Leiden University, Niels Bohrweg 2, 2333 CA Leiden}
\affil[d]{Department of Physics, University of California, Santa Barbara, CA, 93106, USA}

\authorinfo{}

\begin{document} 
\maketitle

\begin{abstract}
Polarization aberrations originating from the telescope and high-contrast imaging instrument optics introduce polarization-dependent speckles and associated errors in the image plane, affecting the measured exoplanet signal. Understanding this effect is critical for future space-based high-contrast imaging instruments that aim to image the Earth analogs with $10^{-10}$ raw contrast and characterize their atmospheres. We present end-to-end modeling of the polarization aberrations for a high-contrast imaging testbed, SCoOB. We use a vector vortex coronagraph (VVC) as the focal plane mask, incorporate polarization filtering, and estimate the peak contrast in the dark hole region 3-10 $\lambda/D$. The dominant polarization aberrations in the system are retardance defocus and tilt due to the OAPs and fold mirrors. Although the mean contrast in the dark hole region remains unaffected by the polarization aberrations, we see brighter speckles limiting the contrast to 1$\times 10^{-9}$ at 1-2 $\lambda/D$. We extend the simulations using the measured retardance maps for the VVC and find that the mean contrast in SCoOB is more sensitive to retardance errors of the VVC and the QWP than the polarization aberrations.  
\end{abstract}

\keywords{Polarization aberrations, high-contrast imaging, Vector Vortex Coronagraphs, space-based telescopes}

\section{INTRODUCTION}
\label{sec:intro}  
High-contrast imaging is a technique to directly image the exoplanets by suppressing the starlight using coronagraphs, wavefront sensing \& control systems. The current spaced-based high-contrast instruments on board HST \cite{debes2019pushing} and JWST \cite{kammerer2022performance} can achieve a planet-to-star contrast $\sim$ 10$^{-5}$-10$^{-6}$, and in the near future, the coronagraph instrument on board the Roman Space telescope aims to achieve a raw contrast of 10$^{-8}$ \cite{poberezhskiy2022roman,poberezhskiy2021roman,kasdin2020nancy}. The Habitable World Observatory (HWO), as recommended by the Astro2020 Decadal survey, consisting of a UV/O/IR telescope, aims to search for habitable biosignatures in planets around many stars with a direct imaging instrument achieving a planet-to-star contrast of 10$^{-10}$.  
\par
Various technologies, such as coronagraphic architectures and wavefront control algorithms used on the space-based high-contrast imaging instruments, are usually tested on high-contrast imaging testbeds (Ex: HCIT, DST \cite{meeker2021twin}, HiCAT \cite{n2013high})  in a vacuum environment that serves as an experimental platform for simulating a space-like environment with dynamic aberrations created by thermal and mechanical vibrations of the telescope and pointing errors. The Space Coronagraph Optical Bench (SCoOB) at the University of Arizona Space Astrophysics Lab is one such testbed using a Vector Vortex Coronagraph (VVC) \cite{mawet2009vector,mawet2010vector,murakami2013design} to achieve $4\times10^{-9}$ contrast at 630nm in 2\% bandpass\cite{vangorkom2022,gorkom_scoob_2024,ashcraft_mueller_scoob_2024}.
\par Among the various aberrations arising from optical surfaces, thermal changes, jitter, and detector noise, high-contrast imaging instruments also suffer from polarization-dependent phase and amplitude aberrations such as retardance piston, defocus, and tilt generated by the telescope and instrument optics and coating. As these low-order aberrations are incoherent, the deformable mirror cannot correct for these aberrations; as a result, they reach the focal plane as ghosts, affecting the achievable contrast \cite{breckinridge2015polarization}. Thus, end-to-end modeling and simulations of the telescope and the instrument using the Polarization Ray Tracing (PRT) algorithm are essential to understand the magnitude of these aberrations and their effect on the contrast, which is routinely done for high-contrast imaging systems \cite{mendillo2019polarization,millar2022polarization,schmid2018sphere,anche2023polarization,anche2023estimation}. The PRT algorithm and description of polarization aberrations has been very well developed in the last two decades and hence will not be described here in detail \cite{chipman2018polarized,breckinridge2015polarization,chipman1995mechanics,breckinridge2018terrestrial} and is also available as an open-source Python package called Poke \cite{ashcraft2022space}.
\par 
In this article, we present the simulations of polarization aberrations for the high-contrast imaging testbed SCoOB using the Zemax OpticStudio API ray trace and performing PRT in
a Python environment and diffraction simulations through a coronagraph using HCiPy\cite{por2018hcipy}. The paper is divided as follows: The optical design of SCoOB is provided in Section \ref{sec:opt-design} for which the Jones and Mueller Pupils are shown in Section \ref{sec:Jones-pupil}. We discuss the retardance and the leakage terms for the Vector Vortex Coronagraph masks and their effect on the contrast for a perfect simulated VVC and the VVC masks
manufactured by Beam Engineering in Section \ref{sec:VVC-masks}. The results are summarized in Section \ref{sec:summary}.

\section{SCoOB optical design}
\label{sec:opt-design}
\begin{figure}[ht!]
    \centering
    \includegraphics[width=0.85\textwidth]{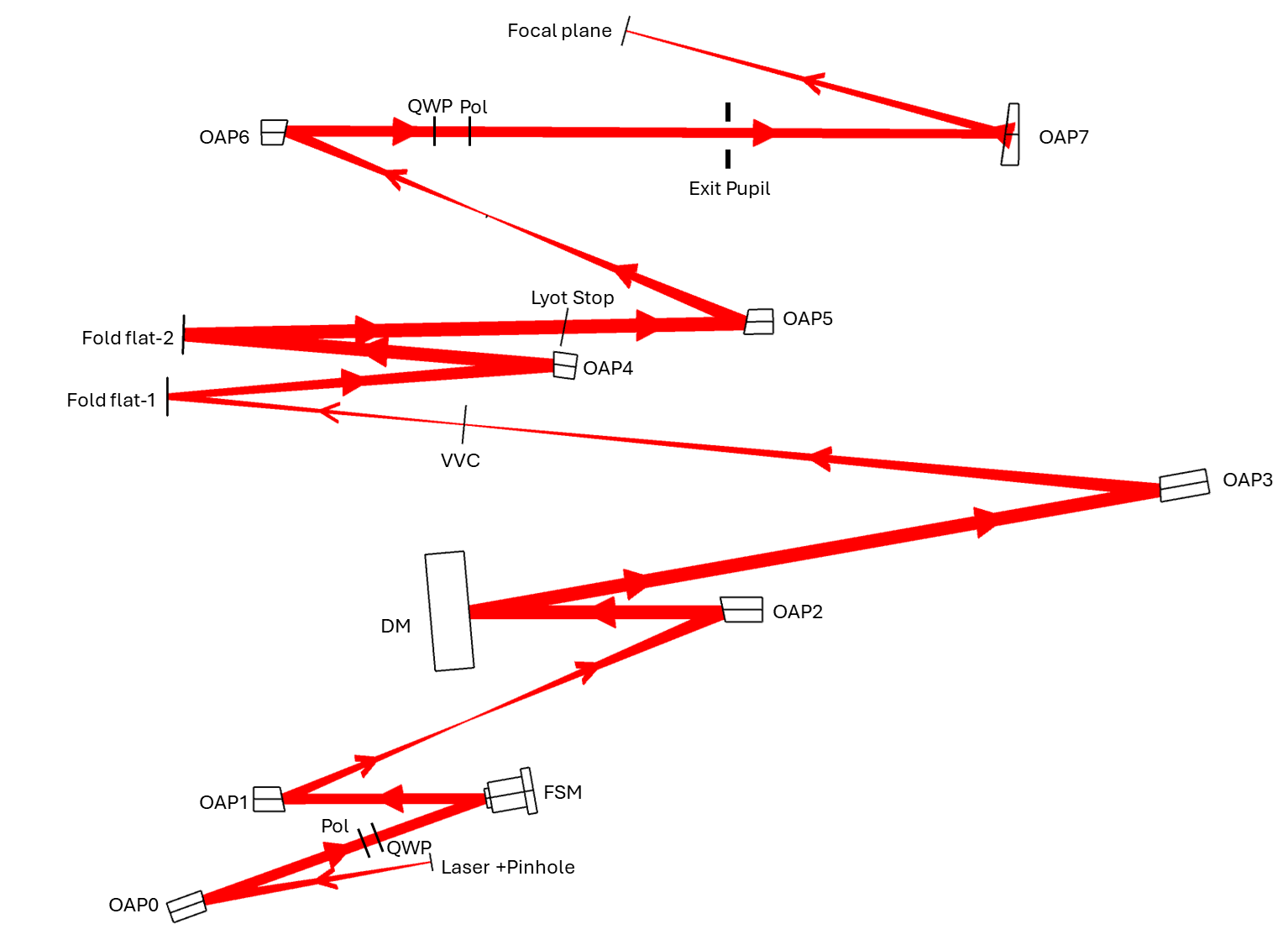}
    \caption{The Optical layout of the Space Coronagraph Optical Bench, which is in the TVAC chamber, consists of eight OAPs, two-fold mirrors, a Fast steering mirror (FSM), a deformable mirror (DM), VVC as the focal plane mask and a Lyot stop. The final beam at the focal plane after OAP7 is F/22. }
    \label{fig:opt-layout}
\end{figure}

The SCoOB optical layout, shown in Figure \ref{fig:opt-layout}, consists of 8 Off-Axis Parabolas (OAP), two folded flat mirrors, a Fast Steering Mirror (FSM), and a Deformable mirror (DM). Detailed information about the optical prescription, wavefront sensing, control algorithms, and dark hole digging procedure can be found in \textit{Ashcraft et al.} \cite{ashcraft2022space} and \textit{Van Gorkom et al.} \cite{vangorkom2022}. SCoOB utilizes a Vector vortex coronagraph (VVC) as the focal plane mask and a Lyot Stop for starlight suppression. The input and output set of crossed circular polarizers are positioned in the collimated space after OAP0 and OAP6 to eliminate the leakage term from the VVC (explained in upcoming sections). The current performance of the testbed is a contrast of 4$\times 10^{-9}$ in a 2\% bandpass and 2$\times 10^{-8}$ in a 10\% bandpass at 630 nm\cite{gorkom_scoob_2024}.

\section{Jones pupil and Mueller pupil}
\label{sec:Jones-pupil}

\begin{figure}[ht!]
    \centering
    \includegraphics[width=1\textwidth]{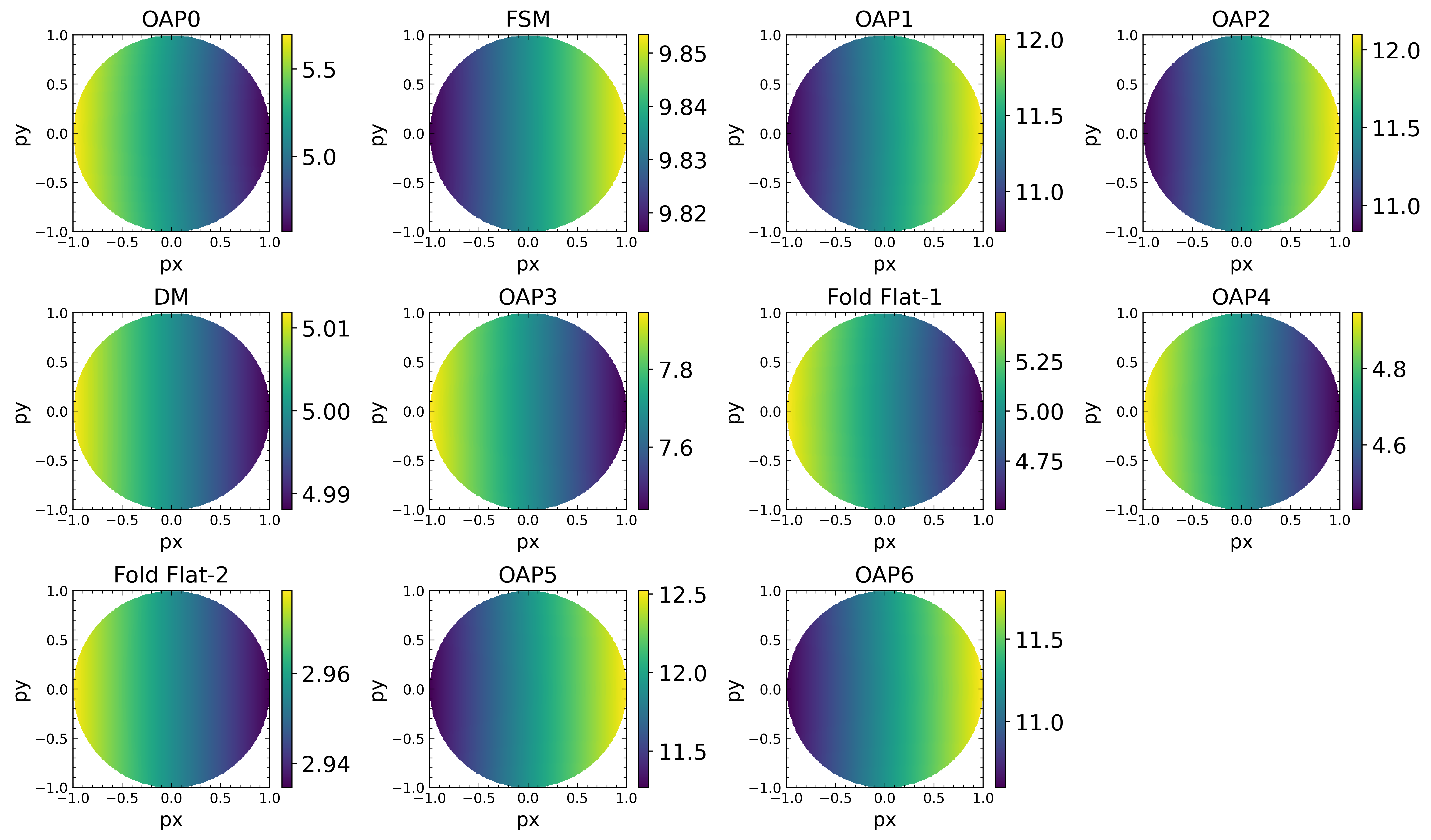}
    \caption{AOI on all the mirrors in SCoOB is shown on the normalized pupil coordinates (px, py).}
    \label{fig:inc-ang}
\end{figure}

We trace a grid of 256$\times$256 rays across the pupil through the optical layout of SCoOB using  Zemax OpticStudio API ray trace. We do not include VVC, Lyot stop, and polarizers in the ray trace, as they would be incorporated later in the HCIPy simulations. The polarization aberrations in any optical system depend on:  1. The radius of curvature/ Angle of Incidence (AOI) 2. coatings on the surface. Using the ray trace, we obtain the AOI on all the mirror surfaces as shown in Figure \ref{fig:inc-ang}. The maximum AOI $\sim$ 12\textdegree~is seen on OAP1 and OAP2, but the AOI varies linearly on the optical surfaces within 1\textdegree~, indicating the dominant polarization aberration will be retardance/diattenuation tilt.  
\begin{table}[ht!]
\begin{center}
\begin{tabular}{ccc}
Surface                     & Avg reflection/transmission & Avg throughput \\ \hline 
7 OAPs (Al+185nm $\rm MgF_2$)      & 0.93                        & 0.6469         \\
Polarizer -Input            & 0.99                        & 0.3202         \\
QWP-Input                   & 0.99                        & 0.3169         \\
FSM (Ag+10nm $\rm Si_3N_4$)         & 0.99                        & 0.3138         \\
DM (Bare Gold)              & 0.93                        & 0.2918         \\
2 Fold Flats (Al+10nm SiO) & 0.95                        & 0.2630          \\
Polarizer -Output           & 0.99                        & 0.2607         \\
QWP-Output                  & 0.99                        & 0.2581    \\ \hline \end{tabular}
\end{center}
\caption{The average reflection/transmission from all the optical components on SCoOB as estimated from the PRT.}
\label{table:reflection_transmission}
\end{table}
\par The coating on all the OAPs is protected Aluminum (Al+185nm MgF$_2$), FSM is protected silver (Ag+10 nm Si$_3$N$_4$), the DM is bare gold, and folded flat mirrors have protected aluminum (Al+10nm SiO). We estimate the Fresnel reflection coefficients and average throughput for all the mirrors in SCoOB as shown in Table \ref{table:reflection_transmission} using the PRT routine in Python. The transmission for the polarizer and QWPs are obtained from the transmission curve from \href{https://www.meadowlark.com/precision-linear-polarizer/}{High Precision Linear Polarizer} and \href{https://boldervision.com/waveplates/aqwp3/} {Achromatic QWP}.
\subsection{Jones pupil}
\begin{figure}[ht!]
    \centering
    \includegraphics[width=1\textwidth]{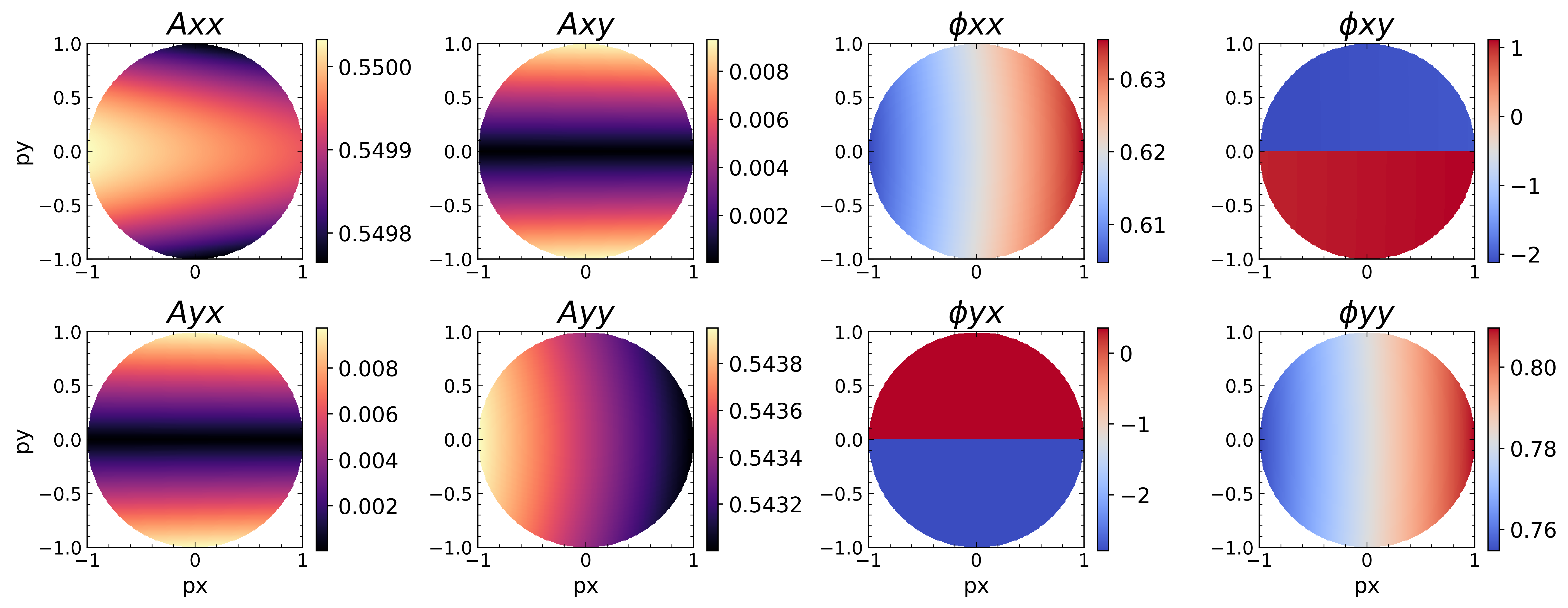}
    \caption{Jones pupil for SCoOB estimated at the exit pupil location (shown in Figure \ref{fig:opt-layout}) for \textit{R} band. \textit{Axx} and \textit{Ayy} show the amplitudes for \textit{X} and \textit{Y} polarized light, respectively, and \textit{Axy} and \textit{Ayx} are the cross-coupled components. Similarly, $\phi xx$ and $\phi yy$ show the phase in radians for $X$ and $Y$ polarized light, and  $\phi xy$ and $\phi yx$ are the cross-coupled components. Although the amplitude terms correspond to the diattenuation aberrations, they do not affect coronagraphic performance. The difference between the $\phi xx$ and $\phi yy$ corresponds to polarization-dependent phase aberrations that cannot be corrected by the DM and thus affect the achievable contrast. }
    \label{fig:jones-pupil}
\end{figure}

The Jones matrices are estimated at the exit pupil at 630nm using the PRT for each input ray with $X$- and $Y-$ polarization as shown in Figure \ref{fig:jones-pupil}. The amplitude \textit{Axx} and \textit{Ayy} vary over the pupil $\sim$ 0.02\% and $\sim$ 0.06\%, the crosstalk components \textit{Axy} and \textit{Ayx} show an amplitude of 0.8\%. 
The phases \textit{$\phi xx$} and \textit{$\phi yy$} are dominated mainly by tilt and piston and vary 0.02 ($\sim$ 2nm) and 0.06 ($\sim$ 6nm) radians over the pupil. The DM will be able to correct for the common aberration between \textit{$\phi xx$} and \textit{$\phi yy$}; hence, the difference of these aberrations are considered in the HCIPy simulations explained in the following section.  
\subsection{Mueller pupil}
\begin{figure}[!h]
    \centering
    \includegraphics[width=1\textwidth]{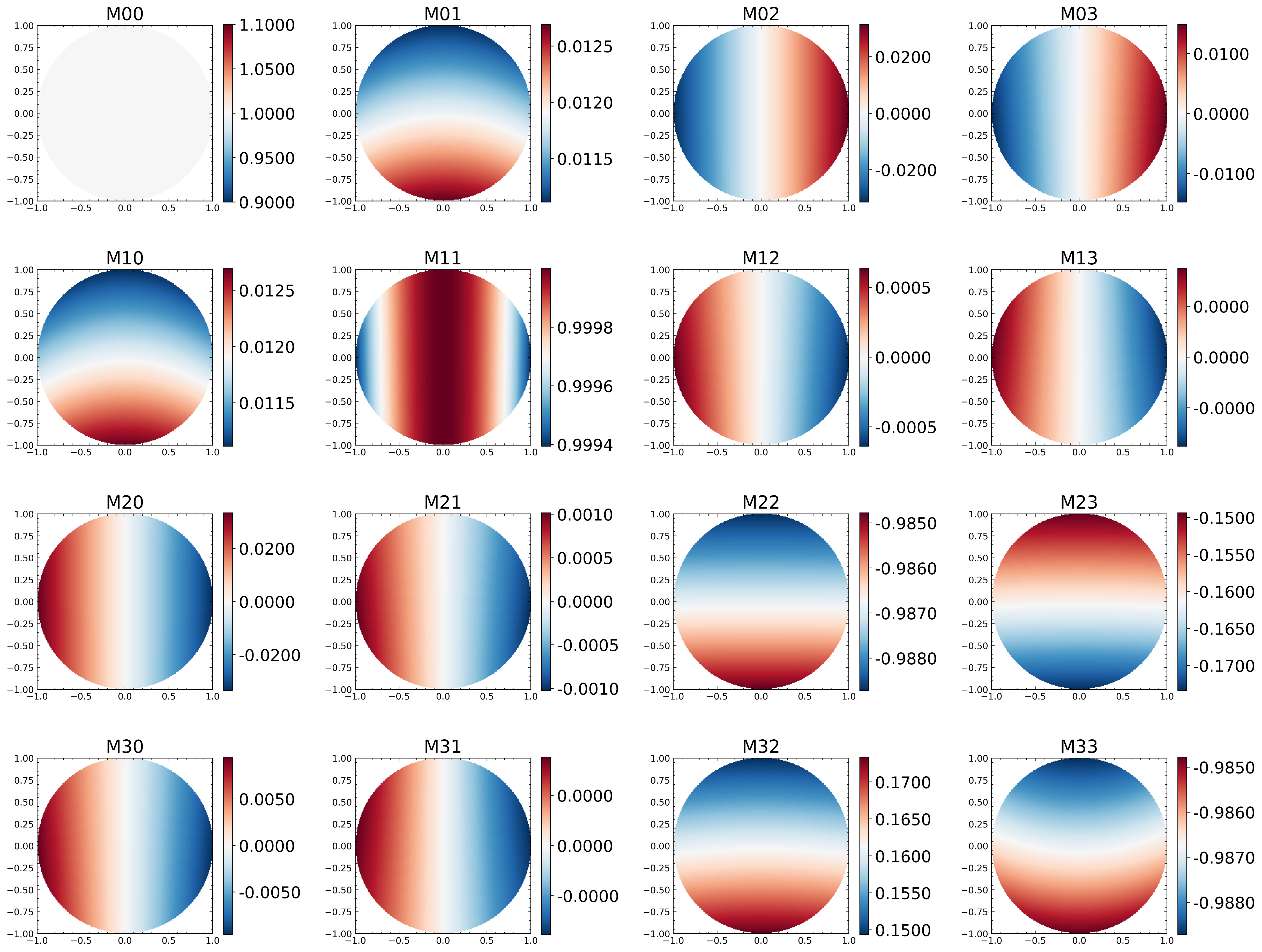}
    \caption{The normalized Mueller pupil for SCoOB obtained from the Jones pupil.The terms M10, M20, and M30 correspond to the linear diattenuation otherwise known as instrumental polarization $\sim$ 2.4\%. The lower 3$\times$3 corresponds to the retardance matrix which shows $<$ 0.05\% of linear retardance and 16\% of crosstalk/circular retardance}
    \label{fig:mueller-pupil}
\end{figure}

The 4$\times$4 Mueller matrix can be estimated by the Kronecker product of the Jones pupils shown in Figure \ref{fig:jones-pupil} \cite{chipman2015polarization,fujiwara2007spectroscopic}. The Mueller matrix for SCoOB at the exit pupil is shown in Figure \ref{fig:mueller-pupil}. All the elements of the Mueller pupil display linear variation due to the dominant aberrations, diattenuation, and retardance tilt from the OAPs and Fold mirrors. The terms M10, M20, and M30 correspond to the linear diattenuation otherwise known as instrumental polarization $\sim$ 2.4\%. The lower 3$\times$3 corresponds to the retardance matrix, which shows $<$ 0.05\%  of linear retardance and 16\%  of crosstalk/circular retardance. When averaged over the pupil, the Mueller matrix will be a diagonal matrix with non-zero M10, M01, M23, and  M32 terms corresponding to the aberrations.   

As we have circular polarizers on the input side, the Mueller matrix is multiplied by the Stokes vector for left circular polarization [1,0,0,1] to estimate the output polarization (columns 1 and 4). As the VVC is a half-wave retarder, the input polarization is flipped to the right circular and propagates through the right circular polarizer (RCP) on the output end, whereas the leakage from the VVC is blocked by the RCP.
\par
To understand the polarization aberrations better, we constructed a polarimeter around SCoOB with a polarization state generator before the input circular polarizers and polarization state analyzer at the exit pupil and measured the Mueller matrix at 630nm. The data were reduced using an open-source polarimetry package called \href{https://katsu.readthedocs.io/en/latest/index.html}{Katsu}. The measurement details and resulting Mueller matrix are provided in \textit{Ashcraft et. al, 2024} \cite{ashcraft_mueller_scoob_2024}. 
\section{Effect on contrast}
\label{sec:VVC-masks}
To estimate the effect of polarization aberrations on the contrast, we perform high-contrast imaging simulations using HCIPy \cite{por2018high}, a physical optics-based simulation framework. In our simulations, we propagate the wavefront defined by the complex electric fields along the optical path of SCoOB. The unpolarized light from the laser is propagated through the input circular polarizers, followed by the Jones pupils up to the VVC, and finally through the output circular polarizers. These simulations are performed for four scenarios: 
\begin{itemize}
\item Perfect VVC + circular polarizers with no polarization aberrations
    \item Perfect VVC  +  circular polarizers with polarization aberrations
    \item Errors on VVC + circular polarizers with polarization aberrations
    \item VVC from BeamCo + errors on circular polarizers with polarization aberrations.
\end{itemize}
\subsection{Perfect VVC and circular polarizers}
\begin{figure}[!h]
    \centering
    \includegraphics[width=1\textwidth]{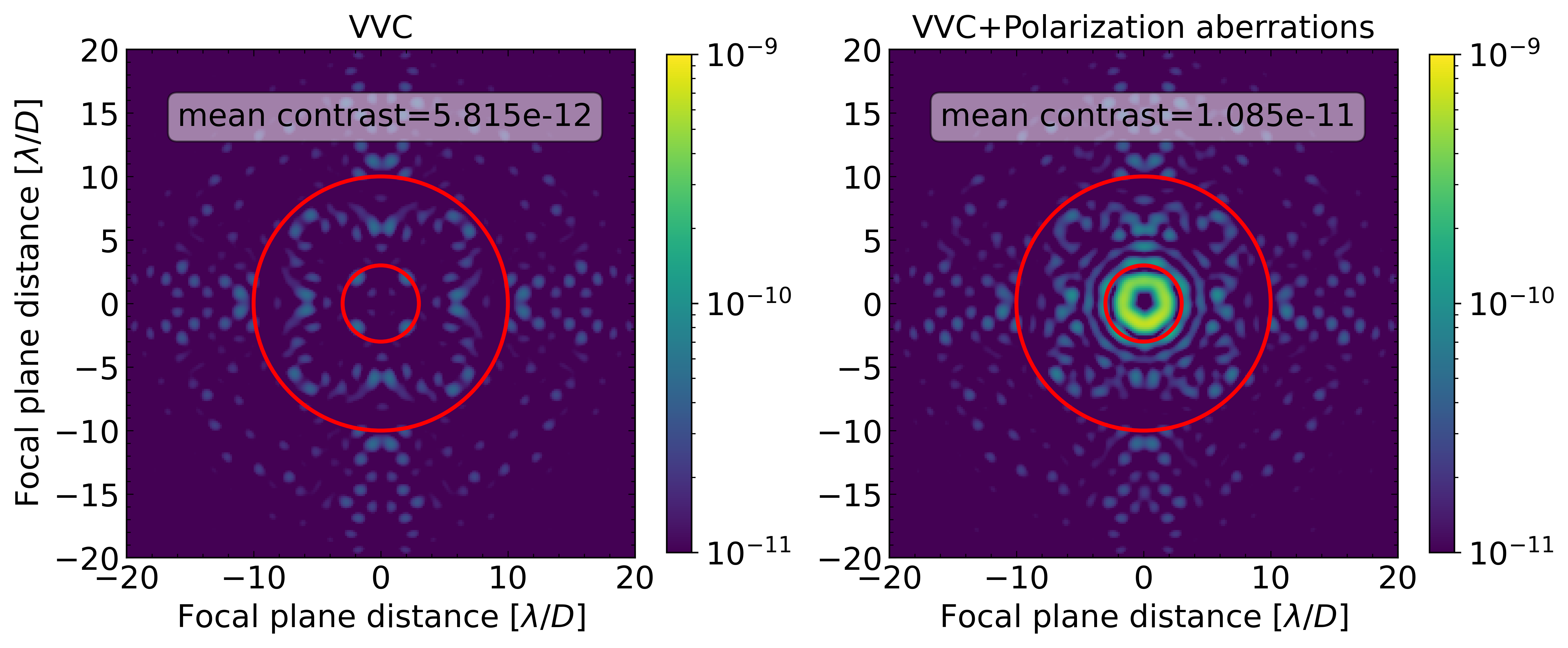}
    \caption{The contrast vs the focal plane distance is shown for a) perfect VVC and circular polarizers without polarization aberrations (left panel) and b) perfect VVC and circular polarizers with polarization aberrations (right panel). The mean contrast estimated in the dark hole region (3-10 $\lambda$/D) is not affected by the polarization aberrations; however, we see brighter speckles $\sim$ 1$\times 10^{-9}$ near 1-3 $\lambda$/D.}
    \label{fig:perfect-vvc}
\end{figure}

In Figure \ref{fig:perfect-vvc}, the contrast versus focal plane distance is shown for the scenario of a perfect VVC and circular polarizers. The left panel represents the case without polarization aberrations (dominated by numerical errors in the representation of the VVC), while the right panel displays the contrast with polarization aberrations from the Jones pupil. The mean contrast in the dark hole region (3-10 $\lambda$/D) is estimated to be 8.54$\times 10^{-12}$ without polarization aberrations and 1.545$\times 10^{-11}$ with polarization aberrations. Although the mean contrast in the dark hole region is well below the expected contrast floor for SCoOB, brighter speckles are observed from the polarization aberrations at inner working angles of approximately 1-2 $\lambda$/D, which limits the contrast to 1$\times 10^{-9}$.

\subsection{Global errors on retardance of QWP, VVC, and leakage from LP}
We extend the high contrast simulations by introducing errors on QWP, VVC, and the leakage from the linear polarizer. The VVC is an achromatic half-wave retarder and is expected to have a retardance of $\pi$ in the desired bandpass. However, it's impossible to manufacture a perfect achromatic half-wave retarder as it's made of many layers of liquid crystal molecules. Similarly, the QWP's retardance and the linear polarizer's extinction ratio may vary due to manufacturing/alignment errors. We incorporate these errors in our simulations by changing the retardance of VVC and QWP by 2-4 degrees and leakage from linear polarizer to be $1\times 10^{-5}$ -  $1 \times 10 ^{-3}$ to estimate their effect on the contrast. The retardance errors on the VVC will result in a leakage that has to be filtered by the circular analyzer after the VVC ( QWP and Pol after OAP6 in Figure \ref{fig:opt-layout}); thus the global retardance errors on the QWPs will result in imperfect filtering by the QWPs affecting the contrast. 
\par
\begin{figure}[!ht]
    \centering
    \includegraphics[width=0.9\textwidth]{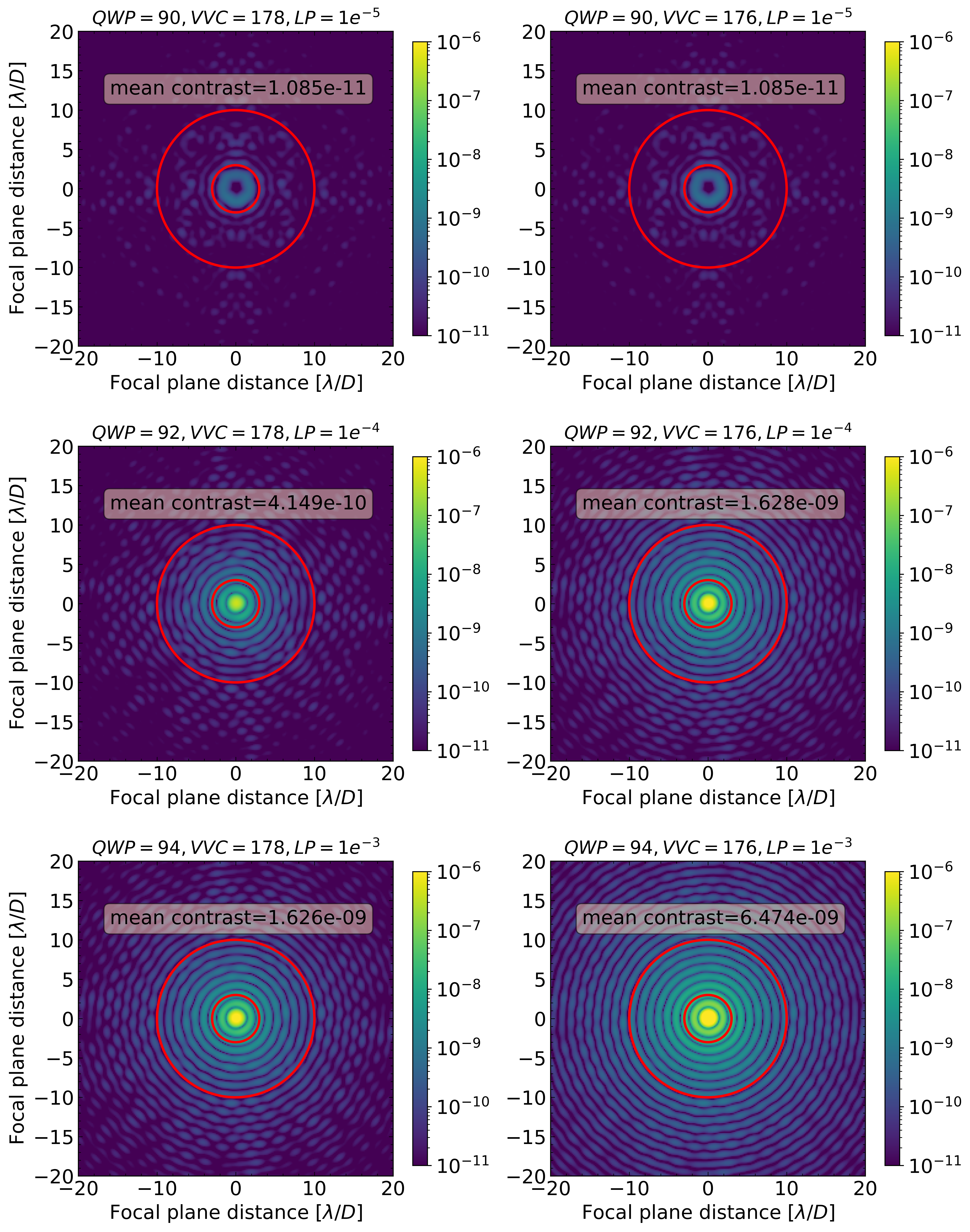}
    \caption{The contrast vs focal plane distance is shown for different cases of global retardance errors on QWP and VVC and leakage from the Linear Polarizer. The retardance errors on the VVC is important only in the case of imperfect filtering by the crossed circular polarizers.}
    \label{fig:imperfect-vvc}
\end{figure}

 Figure \ref{fig:imperfect-vvc} shows the contrast versus focal plane distance for the global retardance errors on the VVC, QWP, and leakage on the LP. The figures on the top panel show that the global retardance error on the VVC does not affect the contrast when there are no retardance errors on QWP and extinction of LP $\sim 10^{-5}$. The figures on the middle panel correspond to the imperfect filtering by the circular polarizers with retardance on the QWP be 92\textdegree changes the contrast by an order of magnitude with global retardance error on VVC increased by 2\textdegree. The speckle residuals increase $\sim$ $10^{-6}$ for IWA $<$ 3$\lambda/D$. The last panel corresponds to the case of QWP retardance of 94\textdegree~, which will result in a mean contrast of ~$10^{-9}$, an order of magnitude lower than the perfect case. The speckle residuals are seen to increase in the dark hole region due to imperfect retardance of VVC and imperfect filtering by the crossed circular polarizers. These simulations show that, to achieve a contrast $10^{=10}$, it is very crucial to have perfect retardance of 90\textdegree for the QWP, a high extinction of linear polarizer, and accurate alignment of the crossed circular polarizers.
\subsection{VVC masks from BeamCo}
\begin{figure}[!ht]
    \centering
    \includegraphics[width=\linewidth]{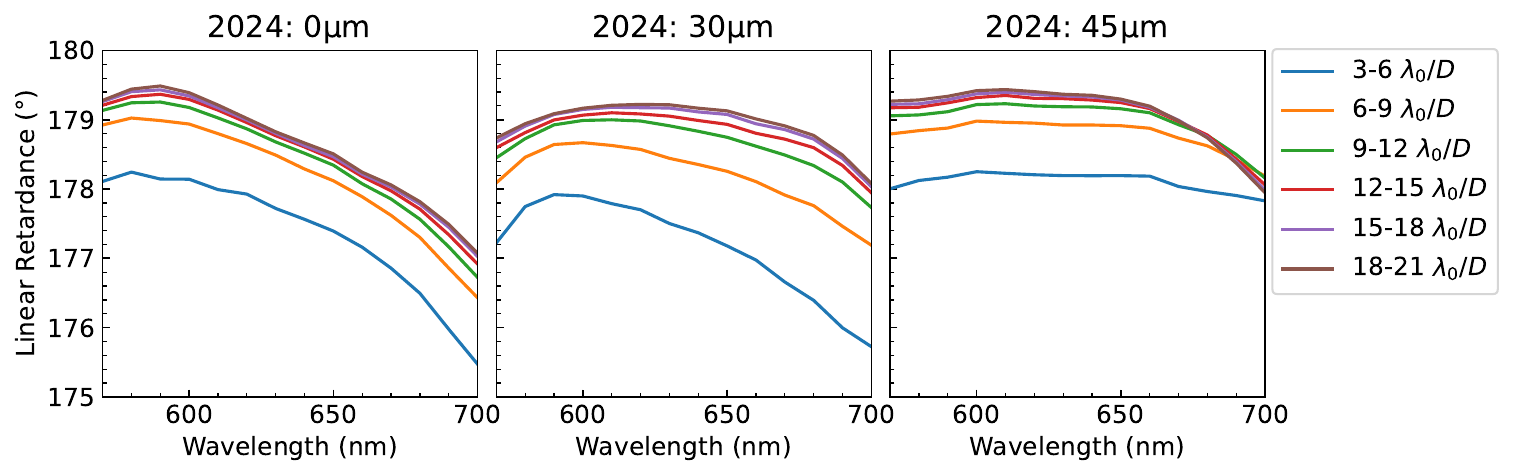}
    \caption{Linear retardance as a function of wavelength for the three Beam Engineering Co. VVC units delivered to UArizona in 2024. The retardance is reported as a mean in binned radial separations in an f/48 beam at $\lambda_0$=635nm. Data from Beam Engineering Co.}
    \label{fig:vvc_units_linret}
\end{figure}

Two VVC focal plane masks were fabricated by Beam Engineering Co. (BEAM Co.) for use in the SCoOB testbed. These are charge-6 VVCs intended to operate over a 20\%  bandwidth centered at 635nm. The VVCs are mounted on Ohara S-BAL35R substrates, manufactured by Rainbow Research Optics, a non-browning radiation-resistant glass with an index of refraction of 1.59. The AR-coated substrates achieve R $<0.5\%$ from 400-700nm. The units were measured by BEAM Co.\ via Mueller matrix microscopy prior to acceptance. The linear retardance within a 1.3mm-diameter centered on the VVC singularity is plotted in Figure \ref{fig:vvc_units_linret}. In the first round of units (2022:0303, 2022:0305, 2022:0404), a 20$\si{\micro \meter}$-diameter opaque dot was placed over the location of the vortex singularity. In the recent round of units, we procured one VVC without an opaque dot (2024:0\si{\micro \meter}) and two with 30 and 45\si{\micro \meter}-diameter dots. 

\par
\begin{figure}[!ht]
    \centering
    \includegraphics[width=\linewidth]{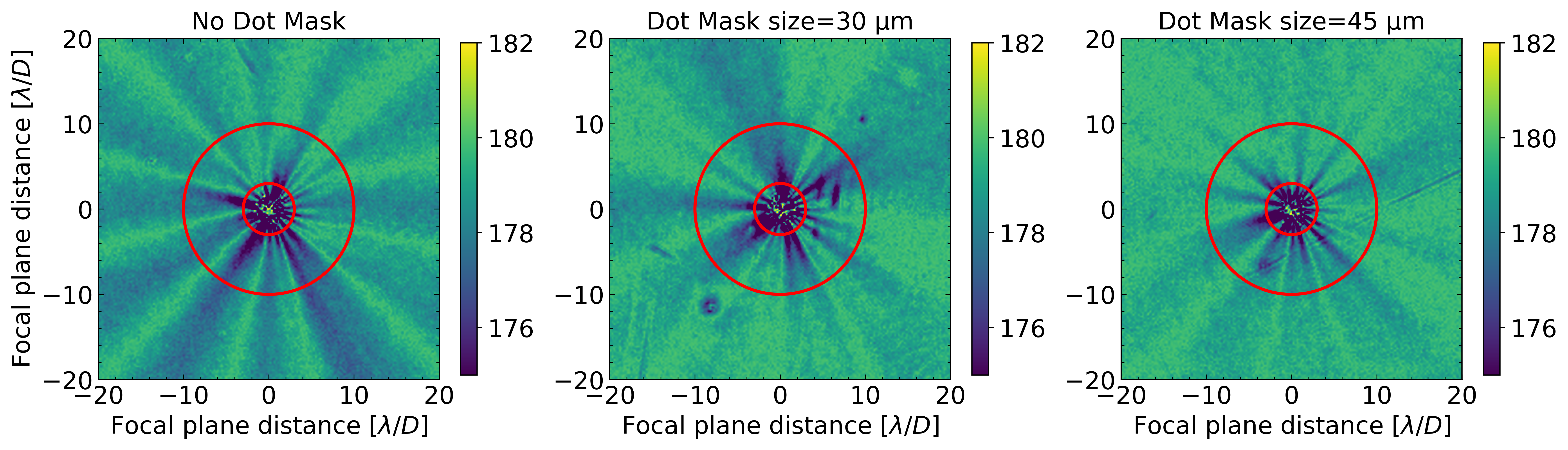}
    \caption{2D Linear retardance maps measured using Mueller matrix microscope by Beam Engineering Co. for the most recent set of VVC masks without and with an opaque dot of diameter 30 and 45 $\mu$m. }
    \label{fig:vvc-beamco}
\end{figure}

\begin{figure}[!ht]
    \centering
    \includegraphics[width=\linewidth]{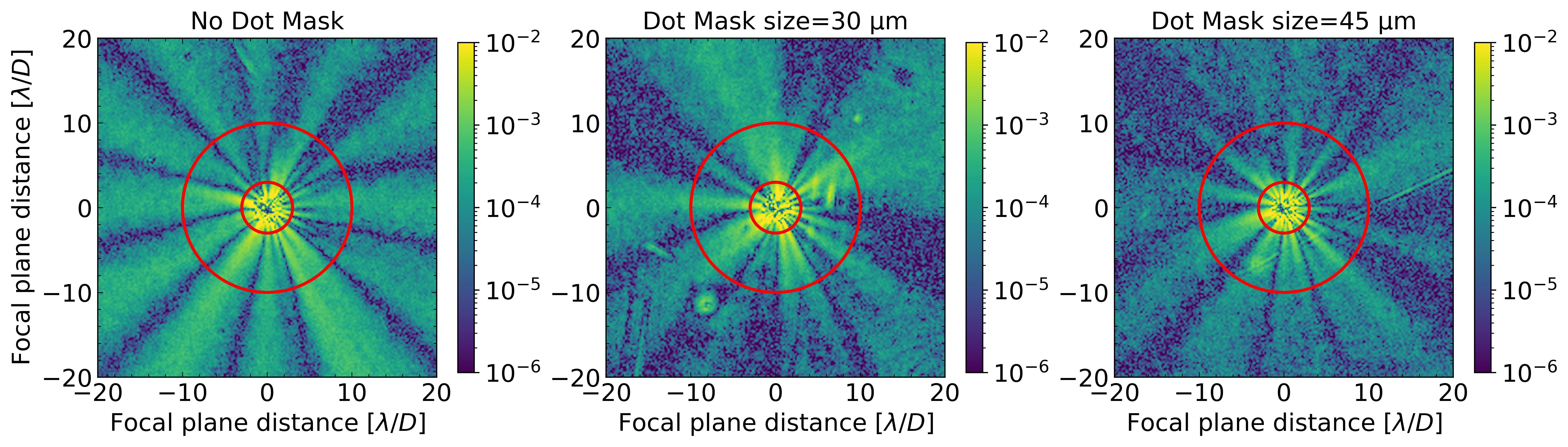}
    \caption{Leakage term originating from the imperfect retardance for the VVCs which gets filtered using the crossed circular polarizers.}
    \label{fig:leakage-vvc}
\end{figure}

 The retardance maps for the recent set of VVCs that were received in 2024 are shown in Figure \ref{fig:vvc-beamco} where the retardance closer to the center in the region $<$ 3$\lambda/D$ is $<$ 176\textdegree ~and increases to 178-180\textdegree~in the dark hole region. The wavelength dependence of the averaged retardance is shown in Figure \ref{fig:vvc_units_linret} at different radial separations for the retardance maps in Figure \ref{fig:vvc-beamco}. The mask with the 45\si{\micro \meter} diameter has the retardance 178-180\textdegree over the wavelength 500-700nm and is expected to have lower leakage and better performance in contrast. The imperfect retardance values on the VVC give rise to a leakage term, ending on the focal plane if not filtered using the crossed circular polarizers. We estimate the leakage term originating from retardance errors for each VVC as explained in \cite{ruane2020experimental} as shown in Figure \ref{fig:leakage-vvc}. The leakage in the region of the dark hole is in the order of $10^{-3}$-$10^{-4}$, which will be filtered using the crossed circular polarizers.
\par
\begin{figure}[!ht]
    \centering
    \includegraphics[width=\linewidth]{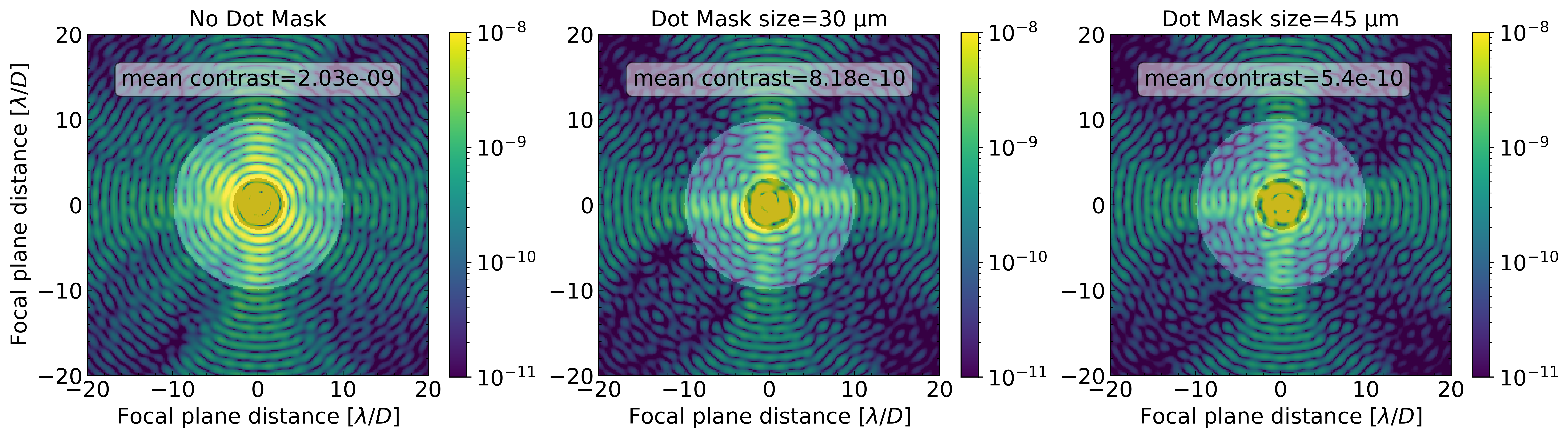}
    \caption{The contrast vs focal plane distance shown for the three VVCs manufactured by Beam Engineering Co. with perfect filtering by the crossed circular polarizers.}
    \label{fig:beamco-vvc-contrast}
\end{figure}

 We then incorporate these VVC masks in our high contrast simulations to estimate the contrast vs focal plane distance for each VVC. As shown in Figure \ref{fig:beamco-vvc-contrast}, we obtain a mean contrast of 2.03$\times 10^{-9}$ for the VVC with no opaque spot, 8.18$\times 10^{-10}$ for the VVC with opaque spot of 30$\mu$m and 5.4$\times 10^{-10}$ for the VVC with opaque spot of 45$\mu$m, respectively. SCoOB has achieved a contrast of 4$\times 10^{-9}$ in 2\% bandpass, comparable to the mean contrast estimated in our simulations.
\section{Summary and conclusions}
Polarization aberrations from the optics of the coronagraph instrument will give rise to polarized speckles in the image plane, affecting the achievable contrast.
\begin{itemize}
    \item We have performed end-to-end polarization aberrations for a coronagraph testbed SCoOB that aims to achieve a contrast better than $10^{-8}$ using VVCs as the focal plane mask.
    \item We performed polarization ray tracing to obtain Jones pupils that show the dominant polarization aberration is retardance tilt. These Jones pupils are converted to Mueller pupils to compare with the measured values.
    \item The polarization aberrations included with an ideal VVC do not limit our mean contrast in the dark hole region; however, they give rise to bright speckles near the IWA $< 3\lambda/D$.
    \item The simulations incorporate global retardance errors on the VVC and QWP and leakage of the linear polarizer to estimate the achievable contrast for the case of an imperfect VVC with imperfect filtering. We find that the mean contrast degrades by an order of magnitude with imperfect filtering, while the retardance errors on the VVC with perfect filtering do not affect the contrast performance.
    \item As a final step, we include the measured retardance maps from the VVCs manufactured by  BEAM Co. and estimate the mean contrast obtained for each of the masks on the order of 1$\times 10^{-9}$.
\end{itemize}
We conclude that these set of simulations demonstrate that polarization aberrations do not affect the contrast performance for SCoOB, which aims to obtain a contrast better than $10^{-8}$ in the dark hole region of 3-10$\lambda/D$; however, it severely degrades the contrast by order of magnitude in the IWA $<$ 3$\lambda/D$ where the future HWO observatory aims to image Earth-like planets directly.
\label{sec:summary}

\appendix    

\acknowledgments 
Portions of this research were supported by funding from the Technology Research Initiative Fund (TRIF) of the Arizona Board of Regents
and by generous anonymous philanthropic donations to the Steward Observatory of the College of Science at the University of Arizona.

\bibliography{references,report} 
\bibliographystyle{spiebib} 

\end{document}